\def\comment#1{}

\newcommand{\beg}{\begin{eqnarray}}
\newcommand{\eee}{\end{eqnarray}}

\documentclass[prl,aps,amsfonts,amssymb,draft,floats,twocolumn]{revtex4}
\def\cm#1{}

\newcommand{\qq}{{\frac{\Psi_1^2}{m_1}}}
\newcommand{\ww}{{\frac{\Psi_2^2}{m_2}}}
\begin{document}
\title{ Phase diagram of  planar  $U(1)\times U(1)$ superconductor
\\  { \small \bf Condensation of vortices with fractional flux    
and a superfluid state}
}
\author{
Egor Babaev 
}
\address{
Institute for Theoretical Physics, Uppsala University 
Box 803, S-75108 Uppsala, Sweden \\
Cornell University 620 Clark Hall Ithaca, NY 14853-2501 \\
Department of Physics, Norwegian University of Science and Technology, N-7491 Trondheim, Norway \\
Science Institute, University of Iceland, Dunhaga 3, 107 Reykjavik, Iceland
}
\comment{\\
}
\begin{abstract}
We discuss a phase 
diagram of  two-dimentional  $U(1)\times U(1)$ superconductor
in the field theoretic formalizm of Ref. \cite{frac}.
In particular we discuss that when penetration length
is short  the  system exhibit a quasi-neutral quasi-superfluid state
which is a state when quasi-long range order sets in 
only in phase difference while individually the phases
are disordered.
\end{abstract}
\maketitle
\newcommand{\la}{\label}
\newcommand{\aaa}{\frac{2 e}{\hbar c}}
\newcommand{\Pfaff}{{\rm\, Pfaff}}
\newcommand{\kA}{{\tilde A}}
\newcommand{\G}{{\cal G}}
\newcommand{\cP}{{\cal P}}
\newcommand{\M}{{\cal M}}
\newcommand{\E}{{\cal E}}
\newcommand{\btd}{{\bigtriangledown}}
\newcommand{\W}{{\cal W}}
\newcommand{\X}{{\cal X}}
\renewcommand{\O}{{\cal O}}
\renewcommand{\d}{{\rm\, d}}
\newcommand{\bfi}{{\bf i}}
\newcommand{\e}{{\rm\, e}}
\newcommand{\bfx}{{\bf \vec x}}
\newcommand{\bfn}{{\bf \vec n}}
\newcommand{\bfE}{{\bf \vec E}}
\newcommand{\bfB}{{\bf \vec B}}
\newcommand{\bfv}{{\bf \vec v}}
\newcommand{\bfU}{{\bf \vec U}}
\newcommand{\ccc}{{\vec{\sf C}}}
\newcommand{\bfp}{{\bf \vec p}}
\newcommand{\f}{\frac}
\newcommand{\bfA}{{\bf \vec A}}
\newcommand{\non}{\nonumber}
\newcommand{\be}{\begin{equation}}
\newcommand{\ee}{\end{equation}}
\newcommand{\ba}{\begin{eqnarray}}
\newcommand{\ea}{\end{eqnarray}}
\newcommand{\bastar}{\begin{eqnarray*}}
\newcommand{\eastar}{\end{eqnarray*}}
\newcommand{\half}{{1 \over 2}}
\section{Introduction}
Berezinskii-Kosterlitz-Thouless  (BKT) transitions 
and phase diagrams of planar superfluids and superconductors
is a subject of wide interest and intensive studies  \cite{1}-\cite{rodr}. 
The present study is motivated by condensed matter systems
with multiple coexistent condensates \cite{mult}-\cite{ashc1}. 
In two-band superconductors a Fermi surface consists of
several disconnected parts which gives rise 
to several types of carriers (which are ``living" on different
parts of the Fermi surface) and correspondingly 
to multiple gaps  \cite{mult}-\cite{we}.
In known two-band superconductors
the two types of carriers are not independently conserved
and the $U(1)\times U(1)$ symmetry 
is softly or strongly broken to $U(1)$.
The two-gap model
with the exact $U(1)\times U(1)$ symmetry  
has been  discussed 
in the theoretical studies  of  liquid metallic hydrogen \cite{ashc1}. 
In such a system (which appears
being close to realization in high pressure experiments)
there should be superconductivity 
of electron-electron and proton-proton Cooper pairs
which gives rise to the $U(1)\times U(1)$ symmetry.
Besides that the two-gap  superconductivity 
is indeed a question of an abstract academic 
interest due to this system albeit 
being formally very simple has deep physics
and exhibit numerous interesting counter-intuitive
phenomena (see also remark \cite{remvol}).
We also mention that certain formal
 aspects of a BKT transition in a square-lattice planar two-gap Abelian
 Higgs model
were earlier also
a subject of study in context of the
lattice  theory which was considered in 
connection with
  spin-charge separated
superconductivity \cite{rodr}. However,
physically,  the phase diagram 
of two-gap system is  principally 
different from the spin-charge separated superconductor.
In the papers by Rodriguez \cite{rodr} it was analyzed a two-component 
model which was suggested  to  be relevant
for spin-charge separated superconductor in a presence of 
a coupling to statistical 
gauge field. The main principal distinction 
of the system \cite{rodr} and a two-gap
superconductor is that in a spin-charge separated
superconductor there exist two vector potentials which
correspond to statistical and ordinary gauge fields
\cite{lee}, \cite{rodr}.
\section{Two-gap Ginzburg-Landau model and
its representation in gauge-invariant variables} 
We begin with a brief outlining of 
basic properties of the  Ginzburg-Landau functional 
for a two-gap   (distinguished  by index $\alpha =1,2$) planar  superconductor:
\beg
F& =&\int d^2x \ \biggl[ \frac{1}{2m_1} \left| \left( \nabla +
i e {\bf A}\right) \Psi_1 \right|^2 + 
\frac{1}{2m_2}  \left| \left( \nabla +
i e {\bf A}\right) \Psi_2 \right|^2 
\nonumber \\
&&+
\eta [\Psi_1^*\Psi_2+\Psi_2^*\Psi_1] 
+ { V} (|\Psi_{1,2}|^2)
+ \frac{{\bf B}^2}{2},
\biggr]
\la{act}
\eee
where $\Psi_\alpha = |\Psi_\alpha|e^{i \phi_\alpha}$ and $
{V} (|\Psi_\alpha|^2)=-b_\alpha|\Psi_\alpha|^2+ 
\frac{c_\alpha}{2}|\Psi_\alpha|^4
$ and $\eta $ is a characteristic of interband Josephson coupling strength
\cite{legg}. In a general case (\ref{act}) may include
other potential terms and mixed gradients terms
which we drop in order have discussion in the simplest form.
 Similar models are also discussed  in particle physics 
\cite{tdlee}.
The situation is quite 
complex in a charged system (\ref{act}) since the 
condensates are  coupled
by the field $\bf  A$. 
However  the variables 
in (\ref{act}) can actually be separated in the London limit
in a simply-connected space.
In \cite{we} it was shown that  the model (\ref{act}) 
is exactly equivalent to an extended version of Faddeev $O(3)$
nonlinear $\sigma$-model \cite{fadde} (which  is also relevant 
in high energy physics \cite{nature} and
for triplet superconductors \cite{tri}).
The version of this model discussed in \cite{we} consists of 
a three-component unit vector $\bfn$ in interaction with a  
vector field $\vec{\sf C}$ and a density-related variable  $\rho$:
\beg
F&=& \frac{\rho^2}{4}(\nabla \bfn)^2 + 
(\nabla \rho)^2 
 + \f{\rho^2}{16}
 \vec{\sf C}^2 + \rho^2 K n_1 +
{{V}}(\rho , n_3 ) 
\nonumber \\
&& + \frac{ 1}{32  }
\left(\nabla_i {\sf C}_j -\nabla_j 
{\sf C}_i -\bfn \cdot \nabla_i
\bfn \times \nabla_j\bfn
\right)^2
\la{e3}
\eee
where $\nabla_i =\f{d}{d x_i}$  and
\beg
 { V}&=& A + B n_3 + C n_3^2 \\  
&& A=\rho^2 [  4c_1m_1^2 + 4c_2m_2^2 -b_1m_1 - b_2 m_2 ] \nonumber \\ 
&&B= \rho^2 [ 8c_2m_2^2 - 8c_1m_1^2 -b_2m_2 + b_1 m_1]  \nonumber \\
&&C= 4\rho^2 [c_1m_1^2 + c_2m_2^2].\nonumber
\eee
A position of the unit vector $\bfn$ on the sphere $S^2$ can be characterized
by two angles as follows:
\be \bfn= (\sin(\theta)\cos(\gamma_n),\sin(\theta)\sin(\gamma_n),\cos(\theta)). \ee
The variables of (\ref{act}) and (\ref{e3}) are related in the following way \cite{we}:
\be  \gamma_n=(\phi_1 -\phi_2); \ee   
\be|\Psi_{1,2}| = 
\left[ \rho \sqrt{2m_1}\  \sin\left(\f{\theta}{2}\right),
\rho \sqrt{2m_2} \  \cos\left(\f{\theta}{2}\right) \right]; \ee 
\beg 
\vec{\sf C } &=&  
\frac{i  }{ m_1 \rho^2}
\left\{\Psi_1^*\nabla \Psi_1-
\Psi_1 \nabla \Psi_1^*\right\}
+\frac{i }{m_2  \rho^2}
\{\Psi_2^*\nabla \Psi_2 \nonumber  \\
&& -\Psi_2 \nabla \Psi_2^*\} 
- \frac{2e}{ \rho^2} \left( \f{|\Psi_1|^2}{m_1} 
+\f{|\Psi_2|^2}{m_2} \right){\bf A}; \ 
\nonumber\la{A}\eee
$ \rho^2 K n_1 = \eta[\Psi_1^*\Psi_2+\Psi_2^*\Psi_1]$,
where  $ K\equiv 2\eta\sqrt{m_1m_2}$ is the Josephson
term. The field $\ccc$ is directly related to supercurrent
\be
\ccc = \f{\bf J}{e \rho^2}
\ee
The mass of the field $\ccc$
is the manifestation of the Meissner effect \cite{we},
the corresponding magnetic field
penetration length is: 
\be
\lambda^2=\frac{1}{ e^2}\left[ \f{|\Psi_1|^2}{m_1}+
\f{|\Psi_2|^2}{m_2}\right]^{-1}=(2 e^2\rho^2)^{-1}.
\ee
The potential term $V$ in (\ref{e3}) 
breaks exact $O(3)$ symmetry associated with $\bfn$
to $O(2)$ and 
defines  the ground state  for $n_3$ (we denote it   by ${\tilde n_3}$), which corresponds to 
uniform density of  condensates  \cite{we}:
\be 
{ \tilde n_3} =\cos \tilde{\theta}= \left[\frac{N_2}{m_2} 
-\frac{ N_1}{m_1}\right]\left[\frac{  N_1}{m_1} + \frac{N_2}{m_2}\right]^{-1}
,\ee
where $N_{1,2}$ stands the for average concentrations
of Cooper pairs $<|\Psi_{1,2}|^2>$. The term

$\rho^2 K n_1$ breaks the remaining $O(2)$ symmetry.
Thus the ground state  of $\bfn$ is a point
where $\cos\gamma_n=-1$ if $\eta >0$
or  $\cos\gamma_n=1$ if $\eta <0$ on 
 a circle specified   
by the condition $n_3={\tilde n}_3$. 
In case when $\eta \approx 0$ the system has
an approximate $U(1)$ symmetry
associated with $\gamma_n$.
 With this we conclude the brief 
reminder of the basic 
 properties of the model (\ref{act}),
more details can be found in  \cite{we,frac}.

Indeed we can study the phase 
diagram of a  two-gap superconductor  
either using  the variables $\bfn$ and $\ccc$ 
or  the  initial variables $|\Psi_\alpha|e^{-i \phi_\alpha}$.
The  former representation is more compact and  convenient,
while   the later variables are more traditional
 so below we discuss the 
phase diagram of the model in terms of both representations.
Additional motivation  for it is that besides
superconductivity the model (\ref{e3}) is of independent
interest being  relevant  in particle physics \cite{nature}. 
\section{Vortices in two-gap superconductor}
In order to understand BKT transitions in a two-gap
system one should understand nature of topological excitations.
The two-gap system allows vortices with fractional
flux \cite{frac} (see also the remark \cite{co5}). In this section
we discuss the physical origin 
of the vortices in the models (\ref{act}) and (\ref{e3}) 
(somewhat extending the discussion in \cite{frac}).
Before we discuss vortices with fractional flux, 
 we would like to remind briefly 
the origin of flux quantization in ordinary superconductors.
First let us however consider a 
topological defect in a {\it neutral} $U(1)$ system like ${}^4$He
which is described by a complex scalar field $|\Psi| e^{i \phi}$
with free energy
\be
F_{neutr}=\f{ 1}{2m} |\nabla \Psi|^2 + a |\Psi|^2 + \f{b}{2} |\Psi|^4 
\ee
Such a system enjoys topological excitations
of the $S^1 \rightarrow S^1$ map (when 
the phase $\phi$ changes $2\pi n$
around the core), in the form of vortices which
have  logarithmically divergent energy per unit length
\cite{pitae}.

Let us now consider a superconductor. 
That is a charged system, so we have a coupling
to the vector potential $\bf A$
\be
F= \f{1}{2m} |(\nabla +i e{\bf A})\Psi|^2 + a |\Psi|^2 + \f{b}{2} |\Psi|^4 +\f{{\bf B}^2}{2} 
\label{**}
\ee
Let us consider a vortex where  the phase changes
$2\pi$ around the core ($2\pi n$
winding of the phase is a topological requirement
which  needed for singlevaluedness of the order parameter).
Then in contrast to neutral case,
away from the vortex core we can minimize
energy by compensating gradients
of phase in the first term of (\ref{**}) by a corresponding
configuration of vector potential {\bf A}, which, as shown
by Abrikosov \cite{aaa}, makes
 this defect being of { finite} energy per unit length.
The corresponding solution for the vector potential
for a vortex characterized by $2\pi$ phase change 
around its core is following:
\be
{\bf A} = \f{{\bf r}\times {\bf e}_z}{|r|}|{\bf A}(r)|
\ee
\beg
|{\bf A}(r)|=\f{1}{er}-
\sqrt{\f{\Psi^2}{m}}K_1\Bigg(e \sqrt{\f{\Psi^2}{m}} r\Bigg)
\label{A}
\eee
where $r$ measures distance from the core 
and ${\bf e}_z$ is a unit vector pointing along the core.

As it was discussed in \cite{aaa}
the above solution for $\bf A$ 
corresponds to the situation 
when a vortex carries a magnetic 
flux quantum $\Phi_0=2\pi/e$. Thus the flux quantization
in the ordinary superconductor
is an effect rooted in topological  and energetical
considerations.

Let us examine now the two-gap model (\ref{act})
in case of $\eta=0$
and show that the two-gap Abelian Higgs model
 can not generate magnetic  flux
which would compensate
gradients of both  order parameters 
in case of a general topological defect.

Let us consider a vortex in (\ref{act}) such that 
 the phase $\phi_1$ has  $2\pi$ winding around core while
$\phi_2$ is constant. Then if a system is ``trying" to compensate
gradient of $\phi_1$ in the first term by a corresponding
configuration of $\bf A$ (that is, by generating flux) at the same
time the field $\bf A$ is inducing  divergence in the second term.
So, by generating exactly one  flux quantum
the system may make the first term being finite but it would
induce a divergence in the second term. On the other hand,
by decreasing flux, the system may minimize divergence
in the second term but then the gradient in the first term
is not compensated by $\bf A$ identically and also becomes divergent.
Thus there are no finite energy
solutions for this type of vortices.
However,  still there is a solution with a minimal energy \cite{frac}  (a reminder
of derivation is given below),
which  indeed should be weakly divergent like in
a neutral $U(1)$ system.
Apparently also in  the case where
the phase $\phi_1$ has $2\pi$
winding around core while $\phi_2$ has a $- 2\pi$ winding,
the system may minimize one of the gradient terms
by generating flux
but at the same time it would increase divergence in the
second term. That is because  to minimize the 
configuration with winding $2\pi$ one should have magnetic
flux going in one direction while for the case $- 2\pi$ in
opposite direction. 
In  the case where the phase $\phi_1$ has
$2\pi n $ winding around core and 
$\phi_2$ also has $2\pi n$
winding the system can generate  flux which
would compensate both gradients \cite{frac}. Only this type of topological
defects in  this system has finite energy.

Let us now describe formally the vortices
with fractional flux following \cite{frac}.
In the London limit ($|\Psi_\alpha|$=const) the eqs. (\ref{act}), (\ref{e3}) become:
\beg
&&F=\frac{\rho^2}{4}(\nabla \bfn)^2 + 
\f{\rho^2}{16}\vec{\sf C}^2 +
\frac{1}{32  e^2}
[\nabla_i {\sf C}_j -\nabla_j 
{\sf C}_i]^2 + \rho^2 Kn_1=
 \nonumber \\
&& \f{ \rho^2}{4} \sin^2\tilde{\theta} (\nabla \gamma_n)^2 +
 \rho^2 \Bigl[ \sin^2\Bigl( \f{\tilde{\theta}}{2}\Bigr) \nabla \phi_1
+\cos^2\Bigl( \f{\tilde{\theta}}{2}\Bigr)\nabla \phi_2 
- e{\bf A}\Bigr]^2
\nonumber \\
&&
+ \f{{\bf B}^2}{2} +\rho^2 K \sin\tilde{\theta}\cos(\phi_1-\phi_2) 
\la{new1}
\eee
Eq. (\ref{new1})  can  also be expressed in traditional variables:
\beg
&&F= 
\f{1}{2}\f{\f{|\Psi_1|^2}{m_1}\f{|\Psi_2|^2}{m_2}}{
 \f{|\Psi_1|^2}{m_1} 
+\f{|\Psi_2|^2}{m_2} 
} (\nabla ({\phi_1-\phi_2}))^2 +
\nonumber \\ 
&&
 \f{2}{  \f{|\Psi_1|^2}{m_1} 
+\f{|\Psi_2|^2}{m_2} }
\Biggl\{\f{|\Psi_1|^2}{2m_1}\nabla \phi_1+
\f{|\Psi_2|^2}{2m_2}\nabla \phi_2 
- e\Biggl( \f{|\Psi_1|^2}{m_1} 
\nonumber \\ 
&&
+\f{|\Psi_2|^2}{m_2} \Biggr){\bf A}\Biggl\}^2+
 \f{{\bf B}^2}{2} 
+ 2\eta|\Psi_1 \Psi_2|\cos(\phi_1-\phi_2)
\la{new01}
\eee
which is a rearrangement of the variables in (\ref{act}) with no apprioximations
involved.
The first term, in the case when $\eta=0$,
describes a {\it neutral} boson (a coupling to $\bf A$
is eliminated when we extract the difference of gauge invariant phase gradients)
associated 
with $\bfn$ (with $O(3)$ symmetry of $\bfn$, in the London
limit being broken down to $O(2)$ by mass terms for $n_3$).
The mass term for $n_1$ breaks the remaining $O(2)$ symmetry.
The second term is the same as the second term 
in (\ref{new1}), together with third term ${\bf B}^2/2=(32  e^2)^{-1}
[\nabla_i {\sf C}_j -\nabla_j 
{\sf C}_i]^2 $
it describes the massive  vector field $\ccc$.

Let us observe that the models (\ref{act})
and (\ref{e3}) have four characteristic length
scales - coherence lengths of condensates,
magnetic field penetration length $\lambda$
and also there is a length scale associated
with intrinsic Josephson effect (which 
is inverse mass for $n_1$). 
Below we discuss 
a system  where both coherence lengths are
of the same order of magnitude and
 where  the characteristic
length scale associated with inverse 
mass for $n_1$ 
is much larger
than magnetic field penetration length $\lambda$
in type-II limit and much larger than
coherence lengths in type-I limit
(which amounts to a negligibly small influence
of the Josephson coupling 
at length scales much smaller than the inverse mass for 
$n_1$).
The  discussion
of the phase diagram of the system 
in the regime of large $\eta$ will
be presented in another publication \cite{moree}. 
We stress that in a  rigorous  sence we should
have $\eta=0$ in order to speak about true BKT transitions,
however when $\eta$ is nonzero but sufficiently small 
we can speak  about some finite-size crossovers.
We also note that in principle it is possible
to have physical systems where $\eta$
is exactly zero - such situation appears
e.g. in theoretical studies of liquid metallic
hydrogen (which appears
being close to being realized in high
pressure experiments) which should allow
superconductivity of protonic and electronic
Cooper pairs \cite{ashc1}. Since electrons can not 
become protons  the condensates
are independently conserved and
we have exact $U(1)\times U(1)$ symmetry.
So in the zero-$\eta$ case, which we  consider in this paper,
the system (\ref{act}), (\ref{e3}) has
a {\it neutral} $U(1)$ symmetry 
associated with the variable $\gamma_n =\phi_1-\phi_2$.

{\it  Vortices in this system,
characterized by the following phase change 
around a core $\Delta \phi_1=2 \pi k_1; \Delta \phi_2=2\pi k_2$,
in what follows we denote for brevity by  $(k_1,k_2)$.}
From (\ref{new1}) and (\ref{new01}) one can observe that 
for  vortices characterized by 
$\Delta( \phi_1 +\phi_2)  \equiv  \oint_\sigma d l [\nabla( \phi_1 +\phi_2)]=4 \pi m;  
\ \  \Delta( \phi_1 - \phi_2)  =0$,
(where we integrate over a closed curve $\sigma$
around a vortex core) the first term in (\ref{new1}) and (\ref{new01})
is identically zero and such vortices is a  analogue of $m$-flux quanta Abrikosov 
vortices in an ordinary superconductor characterized by 
$\f{|\Psi|^2}{m} = \left(\f{|\Psi_1|^2}{m_1} + \f{|\Psi_2|^2}{m_2}\right)$.
It should be observed  that if {\it both} 
 phases $\phi_{1,2}$ change  by $2\pi$ 
around the core then a vortex  carries {\it one}
quantum of magnetic flux.

Let us now  consider the case when $\Delta  \phi_1  = 2\pi$
and $ \Delta \phi_2  = 0$ [a vortex (1,0)].
First of all  such a vortex features  a
 neutral superflow characterized by a $2 \pi$ gain in the variable $\gamma_n$.
As follows from (\ref{new1}), (\ref{new01}),
 the vortex in this respect is equivalent to a vortex in a neutral 
superfluid with superfluid stiffness $ \f{1}{2}\rho^2 \sin^2\tilde{\theta}=
{\f{\Psi_1^2}{m_1}\f{\Psi_2^2}{m_2}}{ \left[
 \f{\Psi_1^2}{m_1} 
+\f{\Psi_2^2}{m_2} 
\right]^{-1}
} $. 
 Besides that, as follows  from (\ref{new1}) and (\ref{new01}),
a topological defect $(1,0)$ is necessarily accompanied by a nontrivial configuration of the 
charged field  $\ccc$. 
That is, for such a topological defect 
the second  term in (\ref{new1})  becomes
$ \rho^2 \left[  
 \sin^2\left( \f{\tilde{\theta}}{2}\right) 
\nabla\phi_1
- e{\bf A}\right]^2$.
Thus, this term is nonvanishing 
 for such a vortex configuration, 
which means that 
this  vortex besides  neutral vorticity  also carries
magnetic field.
Here we stress that a minimal energy solution for 
a vortex with a given winding number is a solution 
where the second  term in (\ref{new1}) is made finite 
by a corresponding configuration of $\bf A$.
Lets us calculate the magnetic flux  carried by such a vortex.
The supercurrent around the core of this vortex 
is:
\be
{\bf J} = 2 e \rho^2 \Big[
\sin^2\Big( \f{\tilde{\theta}}{2}\Big)\nabla \phi_1
- e{\bf A}\Big].
\la{newnew}
\ee
Lets us now integrate this expression over a closed curve $ \sigma $ situated at a
distance  larger than $\lambda$ from the vortex core.
Indeed at a distance much larger than the penetration length the 
supercurrent $\bf J$, or equivalently the massive field $\ccc$, vanishes.
Thus we arrive at the following equation:
\beg
\Phi = 
\sin^2\Big( \f{\tilde{\theta}}{2}\Big) \oint_{\sigma} \nabla\phi_1  d{\bf l} =  
\sin^2\Big( \f{\tilde{\theta}}{2}\Big)  {\Phi_0}
\la{new3}
\eee
where  $\Phi= \oint_{\sigma} {\bf A} d{\bf l} $ is the magnetic flux 
carried by our vortex and $\Phi_0 = 2\pi  /e$ is  the standard magnetic flux quantum.
%
Obtaining equations of motions for $\ccc$ 
from  (\ref{new1})  we can find a solution
for  asymptotic  behavior of the field $\ccc$   for this vortex. At a
 distance larger than coherence length 
$\xi=max [\xi_1, \xi_2]$  from the core the field $\ccc$
behaves as:
\be
|\ccc | = \sin^2\Bigl(\f{\tilde{\theta}}{2}\Bigr)\frac{\Phi_0 }{4 \pi^2 e \rho^2 \lambda^3}K_1\left(\f{r}{\lambda}\right),
\la{cf}
\ee
where $r$ is the  distance from the core.
So this vortex has the following structure:
\begin{itemize}
\item $r<\xi$ vortex core; 
\item $\xi <r < \lambda$ this region features neutral superflow
associated with the gradients of the variable $\gamma_n$ as well as charged
supercurrent associated with the field $\ccc$. 
\item $r>\lambda$ this region features neutral superflow (which
like in a vortex in any neutral system vanishes as $1/r$ away from the core),
whereas at the distance $r>\lambda$ 
the magnetic field and the field $\ccc$ vanish exponentially
according to 
 (\ref{cf}).
\end{itemize}
We can also derive 
the energy per unit length  of this  vortex 
which is: 
\beg
E= E_{c}+ \left[\sin^2\Bigl(\f{\tilde{\theta}}{2}\Bigr)\frac{\Phi_0}{4\pi\lambda}\right]^2
\log{\f{\lambda}{\xi}}
+ \f{\pi}{2}  (\sin\tilde{\theta}  \ \rho)^2
\log\f{R}{\xi},
\la{ener}
\eee where $E_c$ is the core energy and   $R$ is the sample size.
The second term in (\ref{ener}) is the energy of charged current
and magnetic field which 
comes from the second and third terms in  (\ref{new01}).
The third  term in (\ref{ener}) is the kinteic energy of 
neutral current which comes from the first term in  (\ref{new01}).
In an infinite sample, the energy per unit length (\ref{ener}), 
of this vortex in logarithmically divergent
(like in a neutral superfluid)
as  a consequence of the existence of vorticity in the field of
the  massless neutral boson. When $\eta \ne 0$ vortex energy diverges linearly.

We would like to stress that since the two $U(1)$ symmetries
are not independent in this system but are coupled by vector
potential the system actually behaves like a system with 
a $U(1)$ neutral and a $U(1)$ gauged symmetries as it is seen
from separation of variables (\ref{new1}), (\ref{new01}).
The physical origin of the massless neutral $U(1)$ boson
and flux fractionalisation in the two-gap system
can be understood also from the following considerations:
 We can write expressions for individual 
supercurrents in the London limit: 
${\bf j}_\alpha=\f{i e}{m_\alpha}|\Psi_\alpha|^2\nabla \phi_\alpha
 -\f{e^2}{4 m_\alpha}|\Psi_\alpha|^2{\bf A} $, ($\alpha= 1,2$)
from where it  explicitly follows  that { \it even if there are 
gradients of $\phi_1$ while there are no gradients 
of $\phi_2$ nonetheless coupling by $\bf A$ induces
current ${\bf j}_2$ in the condensate $\Psi_2$}. For the 
vortex (1,0) such current partially compensates 
magnetic flux induced by  $\Psi_1$ and 
this compensation leads to existence of 
fractional flux and
an effective neutral superflow
in the system. The fractional flux follows from the condition
that at a distance larger tham $\lambda$  from the vortex core 
the resulting charged current ${\bf J}={\bf j}_1 + {\bf j}_2$ should vanish.
The physical meaning of the neutral superflow is the circulation of
two sorts of charged Cooper pairs  in {\it opposite} directions where
the charged individual currents ${\bf j}_\alpha$ compensate magnetic field induced by  each other.

We also remark that a straightforward inspection of 
(\ref{e3}) shows that this model could allow neutral vortices 
associated with the neutral $O(2)$ boson without a nontrivial 
configuration of the field  $\ccc$ for  any values 
of $\tilde{n}_3$. However
such solutions (e.g. vortices in the case $\tilde{n_3} \ne 0$
characterized
by $\Delta \gamma_n=2 \pi ; \ccc \equiv 0$) are unphysical 
because  these vortices do not satisfy the
condition that $\phi_i$  changes by
$2\pi k_i$ around  the vortex core
(as  follows from (\ref{new1})). 
Thus, while the original model (\ref{act})
and the extended Faddeev model (\ref{e3})
have the same number of degrees of freedom
so that the fields $\bfn$ and $\ccc$
are dynamically independent by construction, 
the mapping incurs a constraint
on topological defects in  $\bfn$ and $\ccc$:
when we go around the vortex 
core  the two phases $\phi_i$ should have $2\pi n_i$ gains 
(with $n_i$ being necessary integer). 

\comment{
So a two-band superconductor (\ref{act})
and extended Faddeev model (\ref{e3}) allow the following linear 
composite topological defects:
\underline{(i) $\Delta (\phi_1-\phi_2)  = 4\pi n$; \ $\Delta(\phi_1+\phi_2) = 0$}:  these are the 
vortices which  feature  neutral superflow. When the ground state of the 
variable $\bfn$ corresponds to the equator on $S^2$  (that is 
the case when $\f{|\Psi_1|^2}{m_1}=\f{|\Psi_2|^2}{m_2} $ or
equivalently  $\cos \tilde{ \theta} = 0 $)
these  vortices do not carry  magnetic flux. In the case 
when $\cos \tilde{ \theta}  \ne 0$, 
the only physical 
solutions with neutral vorticity
also  carry a fraction of magnetic flux quantum. 
 In the case of nonzero Josephson coupling these vortices
are described by the sine-Gordon equation (\ref{new1}).
\underline{(ii) $\Delta \gamma = 0 $; \ $\Delta(\phi_1+\phi_2) = 4\pi k$}
these are the vortices which 
feature circular supercurrent and no circular neutral superflow. Such vortices 
carry $ k $ flux quanta  of magnetic field. 
{ \it One of the  
physical consequences of our analysis is that we can
observe that  in  the
two-gap superconductor  (\ref{act}), (\ref{e3}), 
in spite of the existence of two types of Cooper pairs,
a formation of two sublattices of vortices corresponding
to each of the condensates
in an external field is  
 energetically forbidden. This is because the
energy per unit length  of noncomposite vortices 
is divergent in an infinite sample both in cases of zero and nonzero Josephson
coupling (in  case of finite $\eta$ 
a vortex creates a domain wall
 which makes
its energy per unit length divergent 
in infinite sample \cite{newa,com6}).}
We can also  observe  that   a superconductor 
made up of  one condensate which is of type-II and another 
condensate  which is of type-I will
in general, in the presence of an
external magnetic field,   
form one flux quantum vortices involving both condensates 
and will preserve two-gap  superconductivity, 
even if the external field exceeds the thermodynamic critical 
magnetic field for the type-I condensate \cite{newa}. 
\underline{(iii) $\Delta \gamma = 2\pi n$; \ $\Delta(\phi_1+\phi_2) = 4 \pi l $}:
These are the vortices which   may be viewed
as a co-centered $l$-flux quanta  Abrikosov vortex
combined with  a vortex with neutral vorticity with winding number 
$n$ carrying  a fractional magnetic flux. These vortices
and the vortices of type (ii) characterized by 
$\Delta(\phi_1+\phi_2)  \geq  8 \pi  $ 
should be unstable against decay into more simple vortices
\cite{newa}. }

 
\section{Berezinskii-Kosterlitz-Thouless transitions in two-gap
system}
We begin with a  remark that as it is well known,  
formally, the Abelian Higgs model 
 $H= \f{1}{2m}|(i\nabla - e{\bf A})\Psi )|^2+a\Psi^2 +\f{b}{2}\Psi^4$
does not exhibit
the BKT transition due to interaction between
Abrikosov vortices is of short range, being screened
over the magnetic field penetration length $\lambda$ \cite{1b} (see also a remark \cite{pearl} and a review \cite{1}).
However when the penetration length is significantly large
the system undergoes a ``would be" BKT transition or crossover
(see a review \cite{1})
approximately at the temperature $T^{BKT}=\f{\pi}{2}\f{\Psi^2(T^{BKT})}{m}$.
The equation for $T^{BKT}$  should be solved self-consistently 
with equation for the temperature-dependent gap modulus  $|\Psi(T)|$.
The gap modulus $|\Psi(T)|$
opens at a characteristic  temperature $T^*>T^{BKT}$,
and albeit the phase is random at $T^*>T>T^{BKT}$, however
 $|\Psi(T)|$  has roughly  the meaning of 
the  measure of  the density of  preformed Cooper
pairs in the system \cite{newcom}.
The ``would be" BKT transition in 
a charged system is experimentally observable in extreme
type-II superconductors  \cite{1,orjan3}.
This transition/crossover is associated with 
onset of superconductivity in an extreme type-II
 planar system. In contrast a system with a 
short penetration length does not exhibit planar 
superconductivity at nonzero temperature.

Let us  now discuss the phase diagram
of  planar $U(1)\times U(1)$ superconductor
 in the type-I limit where both coherence 
lengths are larger or equal to  magnetic field
penetration length:
\subsection{ Phase diagram of a planar two-gap system
in the type-I limit}
At finite temperature a planar 
system generates vortices and antivortices.
The BKT transition manifests itself 
as formation of bound vortex-antivortex pairs.
In the regime  $\lambda  \leq  \xi$ we can neglect 
the  presence of the second and third terms in 
(\ref{new01}) while describing
interaction of vortices.
The second and third  terms describe charged supercurrent around 
the vortex core,
 it can be seen from (\ref{new1})
where second and third terms correspond to 
the second and third terms in  (\ref{new01})
 and thus these terms describe  a massive vector field 
$\ccc$. The inverse mass of this 
field is $\lambda$ thus a solution of equations of motion 
for $\ccc$ should be localized at the length scale $\lambda$ and in the 
limit of small $\lambda$ can be neglected
(see also a discussion after eq. (\ref{cf})).
So, when the penetration length 
is very short then we can neglect interaction
of vortices mediated by the charged current
which is exponentially suppressed. 
 On the other hand the neutral
mode [described by the first term in (\ref{new1}), (\ref{new01})]
 is not affected by smallness
of penetration length.
Thus effectively
our system is described by the gradient term of the composite
neutral mode which is (we remind that we consider the case of
the negligible  small $\eta$):
\beg
&&F^{eff  \  (type-I)}= \nonumber \\
&&\f{1}{2}{\f{|\Psi_1|^2}{m_1}\f{|\Psi_2|^2}{m_2}}
 \Biggl[
 \f{|\Psi_1|^2}{m_1} 
+\f{|\Psi_2|^2}{m_2} 
\Biggr]^{-1} (\nabla ({\phi_1-\phi_2}))^2
.\eee
We observe that the 
system has the following vortices with 
{\it identical} configuration 
of the neutral superflow $(1,0)$ and $(0,-1)$
and  antivortices  $(-1,0)$ and $(0,1)$.
A neutral superflow gives rise to a long-range
logarithmic interaction between vortices and antivortices. So,
these vortices undergo a genuine BKT transition 
at the temperature which can be derived immediately
from the first term in 
(\ref{new1}) and (\ref{new01}):
\beg
T_{sf}^{BKT}=\f{\pi}{2}  \rho^2 \sin^2{\tilde\theta}=
\f{\pi}{2} \f{\f{|\Psi_1|^2(T_{sf}^{BKT})}{m_1}
\f{|\Psi^2|_2(T_{sf}^{BKT})}{m_2}}{
\f{|\Psi_1|^2(T_{sf}^{BKT})}{m_1}+\f{|\Psi_2|^2(T_{sf}^{BKT})}{m_2}}
\label{tsf2}
\eee
which, in the case of two-gap
superconductor, should be solved self-consistently with 
the equations for the gap modules $|\Psi_{1,2}(T)|$
\cite{newcom}.
We stress that in a rigorous sence we should
have $\eta=0$ in order to speak about BKT transitions,
when $\eta$ is nonzero then fractional
vortices of the type (1,0) interact linearly
at length scales larger than
inverse mass for $n_1$ 
and thus do not exhibit true BKT transition.
However when $\eta$ is sufficiently small 
we can speak  about some finite-size effects
which, under certain conditions may even be
observable in an experiment. 

We should first emphasis that the transition (\ref{tsf2})
in type-I limit is
associated with formation of vortex-antivortex pairs 
of the following types [(1,0) + (-1,0)],  [(1,0) + (0,1)],
[(-1,0) + (0,-1)], [(0,-1) + (0,1)]
(due to e.g. the vortices (1,0) and (0,-1) have equivalent 
configuration of composite neutral superflow 
in the limit $\lambda \rightarrow \xi$, and thus
in Coulomb gas mapping represent identical charge).
We stress that the interaction of vortices with windings 
of phases belonging to different condensates like e.g. pairs
 [(1,0) + (0,1)] is mediated by neutral superflow 
which has a {\it composite} nature. That is, this
superflow consists of particle current of both
condensates (see a remark in conclusion of the 
previous section).
Besides that the  system allows additional relevant 
composite topological excitation: the one-flux-quantum
vortices (aslo described in \cite{frac})  (1,1) and (-1,-1).
These composite vortices are
 topologically equivalent to the pairs 
 [(1,0) + (0,1)], [(-1,0) + (0,-1)].
{  The composite  vortices  (1,1) and (-1,-1)
play an important role
in this system: these vortices
do not have neutral superflow 
(first term in (\ref{new01}) is identically 
zero for such a vortex configuration)
thus the interaction of these vortices
is exponentially screened, they have finite energy,  and 
remain liberated at any nonzero  temperature.}
Thus, due to existence of the mentioned 
above four equivalent vortex pairing channels 
and the  relevant composite vortices (1,1) and (-1,-1),
 the  phase transition (\ref{tsf2}) marks onset 
of the quasi-long-range order
only in the variable $\gamma_n=\phi_1 -\phi_2$
(described by first term in (\ref{new1}), (\ref{new01})),
while the variable $\gamma_c=\sin^2(\f{\tilde{\theta}}{2})\phi_1+
\cos^2(\f{\tilde{\theta}}{2})\phi_2$
associated with charged boson in (\ref{new1}),(\ref{new01})
remains disordered.
 In the system considered by Rodriguez \cite{rodr}
the counterpart of this transition is a transition 
in the physical electronic phase  (the superconducting transition) thus, the initially 
massless composite mode in \cite{rodr} is
actually coupled to gauge field  like  in an ordinary one gap
superconductor and the phase transition 
in the composite field is transformed in a
``would be" BKT { superconductive}
transition/crossover
 like in a one-gap Abelian Higgs model.
In contrast, we stress the two-gap superconductor
 is a physical
system which possesses a genuine and
potentially experimentally observable massless neutral
boson associated with the composite  phase field $\gamma_n$.
Let us once more stress the meaning of the field $\gamma_n$:
\begin{itemize}
\item
In a two-gap system variables can be separated into a
phase difference and a phase sum. Only the phase sum
is coupled to vector potential.
\item
The gradients of phase sum are gauged
away at the length scale $\lambda$.
(Just like in ordinary Abrikosov vortex a
phase gradient is compensated
by a vector potential at the penetration length scale).
\item  When $\lambda$ is small the long- range
interaction is mediated only by gradients
of phase difference which is not coupled
to vector potential and thus is of long range
and determines the energy of e.g. an (1,0)+(0,1) pair.
When such pairing occurs the individual
phases are indeed disordered [in contrast
to the case if pairing would be only of type (1,0)+(-1,0)
and (0,1)+(0,-1)] and quasi-long-range order sets in only
in phase difference.
Away from a pair (1,0)+(0,1) the 
phase difference is not characterized 
by a nontrivial winding number and its gradients do not 
contribute to kinetic terms of the GL functional. 
At the same time 
the phase sum has a nontrivial winding 
but it is compensated by vector 
potential thus it gives only a finite 
contribution to the GL free energy which 
is not important for the BKT transition (simliar
like a finite energy contribution from vortex cores).
So the system can be mapped onto a Coulomb gas
in a standard way.
\end{itemize}

{ We should stress that in  two-gap
superconductor (\ref{act}) 
the transition in the field $\gamma_n$ has little to do with 
superconductivity since it is a transition in the 
composite neutral field. 
It marks onset of what should be rather called {superfluidity} 
in the field of the composite neutral boson.
Also the important fact is that an elementary topological defect with 
neutral vorticity in the models (\ref{new1}), (\ref{new01})
also carries a fraction of magnetic flux quantum.
Thus albeit we have the phase ordering transition in the neutral
field  however the  fractional flux which  accompanies the  vortices
\cite{frac} allows a detection of  this {\it superfluid}
 phase transition in a two-gap 
superconductor by   standard experimental technique used in studies of
{ \it charged} systems. That is, the vortex-antivortex interaction is mediated 
only  by composite neutral superflow while the magnetic flux 
(at length scale  $\lambda$ from the core) allows observation of these defects 
in  flux-noise measurements or measurements in 
applied dynamic magnetic field.} 
Besides that because of the 
existence of neutral mode, the transition (\ref{tsf2}) 
  in  e.g. flux-noise measurements 
should manifest itself as a sharp well-defined  transition 
which  can not be observed in ordinary charged systems
where one can  see only  a 
``would be" BKT transition, ``washed out" by  Meissner effect.
So,  a two-gap superconductor displays  unique 
potentially experimentally observable novel phase:
in a type-I limit  
 two-gap  superconductor  should exhibit quasi-superfluidity 
at the same time it should { not} exhibit true superconductivity
at any nonzero temperature. 
 In the  type-I limit the true superconductivity
sets in at $T \rightarrow 0$ because at zero temperature
the dimensional reduction 
does not work and the system is effectively 3-dimensional.
In the intermediate case when penetration length 
is  not negligibly small  there could occur a washed out crossover 
to a superconducting phase 
at a nonzero characteristic temperature $T_c$  \ \
($T_{sf}>>T_c>0$). 

Let us now discuss the phase diagram of a two-gap
system in  the case of a large penetration length.
\subsection{ Phase diagram of a planar type-II two-gap system. }
When penetration length is much larger than coherence lengths, the system, again,
 possesses the following vortex excitations:
(1,0), (-1,0), (0,1), (0,-1) 
which still { all} interact with each other
by means of the composite neutral superflow. 
Also there exist composite
vortices without neutral superflow 
(1,1) and (-1,-1). 
{  In contrast to the type-I
limit,  at larger $\lambda$
 the system has {\it two} energetically 
 preferred vortex-antivortex pairing channels 
(instead of four equivalent vortex pairing channels)
associated with binding of the following  pairs [(1,0) + (-1,0)]
and [(0,-1) + (0,1)]. It is  because, as it follows from 
(\ref{new1}) and (\ref{new01})
the vortices  (1,0) and (-1,0) have attractive interaction
{\it both} due to neutral superflow and charged current, in contrast
the vortices  (1,0) and (0,-1)  interact
only due to their neutral modes [which is a composite
mode, whereas  the 
charged supercurrent which accompanies
 the vortices  (1,0) and (0,-1) consists 
only of Cooper pairs of sort ``1" in the case
(1,0) and sort ``2" in the case (0,-1)  (see  the  concluding remark
 in section III)].}

Let us  illustrate the discussion 
with the vortex (1,0).
The free energy density  for this configuration becomes:
\beg
F&=& 
\f{1}{2}\f{|\Psi_1|^2}{m_1}\f{|\Psi_2|^2}{m_2}\left[
 \f{|\Psi_1|^2}{m_1} 
+\f{|\Psi_2|^2}{m_2} \right]^{-1}
 (\nabla \phi_1)^2 +
 \nonumber \\ & & 
{2}\left[  \f{|\Psi_1|^2}{m_1} 
+\f{|\Psi_2|^2}{m_2} \right]^{-1}
\Biggl\{\f{|\Psi_1|^2}{2m_1}\nabla \phi_1 
- 2e{\bf A}\Biggl\}^2
 \nonumber \\ & &
+ \f{{\bf B}^2}{2}.
\la{ne01} \eee
From this expression it is seen that 
the interaction of  vortices (1,0) and (-1,0) 
at the length scales smaller than $\lambda$ is characterized by the 
effective stiffness
\be
 J^{eff}_1=J^n + J^c_1, \ \ \ \ \ {\rm where}  \
J^n(T)=\f{\f{|\Psi_1|^2(T)}{m_1}\f{|\Psi_2|^2(T)}{m_2}}{
 \f{|\Psi_1^2|(T)}{m_1} 
+\f{|\Psi_2^2|(T)}{m_2} 
}
\label{je1}
\ee
\be
{\rm and} \ J^c_1(T)=\f{\f{|\Psi_1|^4(T)}{m_1^2}}
{  \f{|\Psi_1|^2(T)}{m_1} 
+\f{|\Psi_2|^2(T)}{m_2}}
- \left({ corrections \ due  \atop \ to \  finite \ \lambda}\right)
\label{jc1}
\ee
The contribution to the interaction between these vortices
which is mediated by charged current (characterized 
by the stiffness $J^c$) should receive
corrections due to charged current vanishes
over the magnetic field penetration length $\lambda$.
Similarly, the effective stiffness for interaction between vortices
(0,1) and (0,-1) is given
 by an analogous expression
\be
J^{eff}_2=J^n + J^c_2.
\label{je2}
\ee
where 
\be
J^c_2(T)=\f{\f{\Psi_2^4(T)}{m_2^2}}
{  \f{\Psi_1^2(T)}{m_1} 
+\f{\Psi_2^2(T)}{m_2} } 
- \left({  corrections \ due  \atop \ to \  finite \ \lambda}\right)
\label{jc2}
\ee

Thus the system in the regime 
$\lambda \to \infty$ will have two transitions 
at the following temperatures:
\beg
T^{BKT}_1=\f{\pi}{2} J^{eff}_1(T^{BKT}_1) \nonumber \\  \ T^{BKT}_2=\f{\pi}{2} J^{eff}_2(T^{BKT}_2)
\label{ttt}
\eee
This is a consequence of the fact that making $\lambda$ 
large, reduces four vortex pairing channels
(which are equivalent 
in the type-I limit)
to two non-equivalent energetically preferred pairing channels.
We should emphasis that the 
 composite co-centered excitations
(1,1) and (-1,-1) which do not carry 
superflow   [and which are
partially  responsible
for disordering the variable  $\gamma_c=\sin^2(\f{\tilde{\theta}}{2})\phi_1+
\cos^2(\f{\tilde{\theta}}{2})\phi_2$ in the 
short penetration length limit] are irrelevant 
if the penetration length is large since the effective stiffness
of interaction between vortices (1,1) and (-1,-1)
is $J= [|\Psi_1|^2/m_1+|\Psi_2|^2/m_2]$
(as it follows from (\ref{new01}), (\ref{new1}))
and thus when we can not assume that interaction 
is completely screened due to Meissner effect, then the large
effective stiffness which characterizes 
interaction of these defects 
prevents existence of  liberated vortices 
of this type at the temperatures 
(\ref{ttt}).
We also remark that in the type-II
limit the interaction of vortex pairs
of the type  $(1,0)+(0,1)$
is getting depleted at the length scales
$r<\lambda$ as shown in appendix A.

\comment{
From the form of the effective stiffnesses $J^{eff}_{1,2}$
we can understand  the experimental manifestation 
of these transitions: let us assume that $T_1^{BKT} > T_2^{BKT}$
and $|\Psi_1|^2(T)/m_1 >> |\Psi_2|^2(T)/m_2$.
The interaction between vortices
which governs  both
transitions  consists of contributions
of  neutral and charged currents (except for the case 
when one of the condensates reaches critical stiffness
of the BKT transitions at a temperature higher than 
the temperature when the second gap opens). 
Then from  (\ref{je1}) and   (\ref{jc1})  we can  
observe that in the situation when
$|\Psi_1|^2(T_1^{BKT})/m_1 >> |\Psi_2|^2(T_2^{BKT})/m_2$
the interaction  which governs binding transition
in the system of vortices (1,0) and (-1,0)
is dominated by charged supercurrent (at temperatures around $T^{BKT}_1$).
It is because the charged current,  in this
situation, has larger stiffness $J^n(T^{BKT}_1 ) << J^c_1(T^{BKT}_1 )$.
 At the same time
the second transition [associated with binding of the
vortices (0,1) and (0,-1)] is predominantly mediated
by the neutral mode $(J^n(T^{BKT}_2 ) >> J^c_2(T^{BKT}_2 ))$.
Thus in a such a system the BKT transition, which is associated
with establishment of quasi-long-range order
in the  condensate with higher stiffness   $|\Psi_1|^2/m_1$, should manifest itself
as a washed out crossover like in a charged system.
At the same time the lower transition at $T^{BKT}_2$
should have features of a sharper transition 
like in a neutral system without Meissner screening
(for a discussion of degradation of BKT transition due to
Meissner screening see \cite{1}).
The character of experimental manifestation 
of these transitions should also 
depend on the following circumstance:
in the case when 
$|\Psi_1|^2(T^{BKT}_2)/m_1$ is much larger than 
$|\Psi_2|^2(T^{BKT}_2)/m_2$
the fraction of flux quantum confined by 
the vortices
$(0,\pm 1)$ will be very small
and this may make it particularly
hard  to detect the transition the system
of vortices $(0,\pm 1)$ in  flux noise measurements or
measurements in applied dynamic magnetic field.}

\section{Conclusion}
We presented a
discussion of  the  phase diagrams of 
 planar $U(1)\times U(1)$ superconductors and 
predicted what could be a 
novel quasi-superfluid state realized in two-gap
superconductor 
in a type-I limit. 
The physical meaning of the quasi-superfluid state in a
two-component charged system is a quasi-long range 
order in phase difference when individually the phases
are disordered. Such a state leads to a counter-intuitive
situation of purely topological origin
when there are no disspationless individual currents
while a superflow of oppositely directed two
charged currents is dissipationless. 
%


We also stress that the present discussion is 
relevant for much wider class of physical models
than two-gap superconductivity:
the models (\ref{act}), (\ref{e3}) are also  relevant 
 in the high energy physics \cite{tdlee,nature}.

We are grateful  to  L.D. Faddeev, A. J. Niemi, V. Cheianov,
O. Festin, S. Girvin, A. Gurevich,  A.J. Leggett, P. Svedlindh, A. Sudbo, D. Agterberg, 
F. Wilczek and especially to 
 G.E. Volovik
and K. Zarembo  for many discussions and/or comments. We thank 
J. Rodriguez for informing us about the studies \cite{rodr}
and useful comments. 
This  work  has
been supported by grant STINT IG2001-062, the Swedish
Royal Academy of Science and Goran Gustafsson Stiftelse UU/KTH,
and the Research Council of Norway, Grant No. 157798/432.
We acknowledge the support of the Bergen Computational Physics
Laboratory in the framework of the European Community - Access
to Research Infrastracture action of the Improving Human Potential Programme.
The author also acknowledges NorFA mobility scholarship and 
Thordur Jonsson and Larus Thorlacius for hospitality 
during his stay at University of Iceland.

{\bf A note added in proof:} We call the reader's  attention 
to a problem of a BKT transition similar to the  discussed above type-II 
limit, which appears in high-$T_c$ superconductors
and was considered in \cite{bl} and references cited therein.
We thank G. Blatter for informing us about these references.
Another, related from a field theoretic point of view,
problem was considered in \cite{wil}.

\section{Appendix A}
The pairing of vortices and antivortices of the type $(1,0)+(0,1)$
is possible because two condensates  are coupled by vector
potential so gauge invariant equations for currents
of two types of Cooper pairs are:
${\bf j}_\alpha=\f{i e}{m_\alpha}|\Psi_\alpha|^2\nabla \phi_\alpha
 -\f{e^2}{4 m_\alpha}|\Psi_\alpha|^2{\bf A} $, ($\alpha= 1,2$)
so that  if there are 
gradients of $\phi_1$ while there are no gradients 
of $\phi_2$ nonetheless coupling by $\bf A$ induces
current ${\bf j}_2$ in the condensate $\Psi_2$. For this reason a
vortex $(1,0)$  interacts with a
a vortex  with phase winding in different condensate:
$(0,1)$. 
Let us now consider a vortex $(1,0)$.
It has the following particle currents of condensates 
$\Psi_{1,2}$ around its core:
\beg
{\bf j}_1=\f{i e}{m_1}|\Psi_1|^2\nabla \phi_1  -\f{e^2}{4 m_\alpha}|\Psi_1|^2{\bf A}, \nonumber \\
{\rm and} \  \  \  \  \  \  \  \  \  {\bf j}_2=-\f{e^2}{4 m_\alpha}|\Psi_2|^2{\bf A}
\label{currents}
\eee
The standard solution for the vector potential of an Abrikosov vortex is:
\be
{\bf A} = \f{{\bf r}\times {\bf e}_z}{|r|}|{\bf A}(r)|
\ee
where $r$ measures distance from the core 
and ${\bf e}_z$ is a unit vector pointing along the core.
For such a vortex the solution for vector potential is 
\beg
|{\bf A}|=\f{\qq}{\qq+\ww}\f{1}{er}-
\f{\qq}{\sqrt{\qq+\ww}}K_1\Bigg(e \sqrt{\qq+\ww} r\Bigg)
\label{A}
\eee
For $r << \lambda = \Big(e \sqrt{\qq+\ww}\Big)^{-1}$
eq (\ref{A}) reads
\be
|{\bf A}| \approx \f{\qq}{\sqrt{\qq+\ww}} \left[
\f{1}{2} {\rm ln}\left({\f{r}{2\lambda}} \right) - \f{1  - 2\,\gamma}{4} \right] \f{r}{\lambda}
\ee
For this reason for the mentioned above
vortex pair, in  a type-II superconductor with the
separation $r$ when
$\lambda  >>  r >>  {\rm max}_i [\xi_i (T_{\rm \scriptscriptstyle BKT})] $,
the interaction mediated by
the neutral mode due to coupling
by vector potential,  which is composed of oppositely circulating 
supercurrents of two species of Cooper pairs,  is depleted because  
a vortex $(1,0)$
at length scale  $r<<\lambda$
has depleted current ${\bf j}_2$
likewise a vortex $(0,1)$
also has depleted current ${\bf j}_1$.


\begin{thebibliography}{99}
\bibitem{newcom}
For a recent review of similar 
calculations in different models,
including superconductivity 
and particle physics-related models  see \cite{2}.
Particularly deep insight into this scenario 
can be gained from the lattice study \cite{costas}
and comparing it with studies of 
similar models with $Z_2$ symmetry \cite{costas2}.
For earlier discussions 
of this scenario in High-Tc superconductors
see \cite{1c} and references cited therein.
\end{thebibliography}
\end{document}